\documentstyle[12pt]{article}
\begin{document}
\begin{center}
{\large PECULIARITIES OF MATTER MOTION IN METRIC-AFFINE GRAVITATIONAL THEORY}\\
\vskip 0.4cm
BABOUROVA\enspace O.V., FROLOV\enspace B.N. and KOROLIOV\enspace M.Yu. \\
\vskip 0.4cm
{\it Department of Mathematics, Moscow State Pedagogical University,\\
     Krasnoprudnaya 14, Moscow 107140, Russia;\\
     E-mail: baburova.physics@mpgu.msk.su, frolovbn.physics@mpgu.msk.su}
\vskip 1cm
\end{center}

\begin{abstract}
{\small
\par On the basis of the Lie derivative method in  a  metric-affine
space-time $(L_{4},g)$  it is shown  that  in   the   metric-affine
gravitational  theory  the  energy-momentum  conservation  law  and
therefore the equations of the matter motion  are  the  consequence
(as  in  the GR)  of  the  gravitational  field   equations.   The
possibility  of  the  detection  of   the   space-time   non-metric
properties is discussed.}
\end{abstract}
{\bf 1. Introduction}
\vskip 0.2cm
\par
It is well known that in the General Relativity the  equations
of the matter motion are the consequence of the gravitational field
equations. The reason  of  this  fact  consists  in  the  identical
vanishing (as the cosequence of  the  Bianchi  identities)  of  the
covariant divergence of the left part of  the  Einstein  equations,
which leads to the covariant energy-momentum  conservation  law  of
matter
\par
\begin{equation}
\stackrel{(R)}{\nabla}_{\beta }T_{\alpha }\!^{\beta} = 0  \label {eq:1}
\end{equation}
(here $\stackrel{(R)}{\nabla}$ is the covariant differentiation  with
respect  to  the  connection of a Riemann space-time). The equations
of  the  matter  motion are the consequence of (\ref{eq:1}).
\par
In the Einstein-Cartan theory \cite{Kib,Tr} the same  situation  takes
place \cite {He:let}. The more complicated case occurs in the generalized
theories of gravity with torsion in a Riemann-Cartan space-time $U_{4}$
based on  non-linear  Lagrangians,  where  the  some  new  covariant
identities appear. In the theories of a such type as it  was  shown
in \cite{Bab:Pon,Bab:dr} the equations of the matter motion are also the
consequence of the gravitational field equations.
\par
In this paper we generalize the results of \cite{Bab:Pon,Bab:dr} to the
theory of gravitation in the  metric-affine  space-time $(L_{4},g)$  with
the curvature $R_{\mu \nu \sigma}\!^{\lambda }$, torsion $T_{\mu \nu}\!
^{\lambda}$ and nonmetricity $Q_{\lambda }\!^{\sigma \rho }:= \nabla
_{\lambda }g^{\sigma \rho }$. The metric-affine theory of gravitation was
proposed in \cite{He:nat} and now is of a great interest in connection
with the problem of the relation of the gravitation and the elementary
particles physics \cite{Ne:Si,He:Si}.
\vskip 2cm
\noindent
{\bf 2. The energy-momentum conservation law as the consequence of
the field equations in the metric-affine gravitational theory}
\vskip 0.2cm
\par
We shall use  the  Lie  derivative  method  for  deriving  the
differential identities in the metric-affine  space $(L_{4},g)$.  In  a
Riemann-Cartan space this method was described in  detail in \cite{Kop}.
Let us consider the  Lie  transport $\pounds_{\xi }$ in the direction of an
arbitrary vector field $\xi ^{\sigma }$. Then the transformation law of the
gravitational field Lagrangian density ${\cal L}_{0} = \sqrt{-g}L_{0}$ is
\begin{equation}
\pounds_{\xi }{\cal L}_{0} = \xi ^{\sigma }\nabla _{\sigma }{\cal L}_{0} +
{\cal L}_{0}\hat {\nabla}_{\sigma }\xi ^{\sigma }\; , \label{eq:2}
\end{equation}
where $\hat {\nabla}_{\sigma }:= \nabla _{\sigma } + T_{\sigma }$,
and $T_{\sigma }:= T_{\sigma \tau }\!^{\tau }$. Here $\nabla $  is  the
covariant differentiation with respect to the connection of $(L_{4},g)$.
\par
{}From the other side the result of this Lie differentiation can
be calculated as the consequence of the explicit  dependence of the
Lagrangian density
\begin{equation}
{\cal L}_{0} = \sqrt{-g}L_{0}(g^{\sigma \rho },\;
R_{\mu \nu \sigma }\!^{\lambda },\; T_{\mu \nu }\!^{\lambda }) \label{eq:3}
\end{equation}
from the metric, curvature and torsion tensors of $(L_{4},g)$, the Lie
differentiation of those being  calculated
by the corresponding rules that valid in $(L_{4},g)$:
\begin{eqnarray}
\pounds_{\xi }g^{\sigma \rho } = \xi ^{\mu }Q_{\mu }\!^{\sigma \rho }
-  2\xi ^{(\sigma \rho )}\;,\;\;\;\;\;
\xi_{\mu}\!^{\rho}:= \nabla _{\mu }\xi ^{\rho } -
T_{\mu \nu }\!^{\rho }\xi ^{\nu }\; , \label{eq:4} \\
\pounds_{\xi }R_{\mu \nu \sigma }\!^{\lambda } = \xi ^{\rho }\nabla _{\rho }
R_{\mu \nu \sigma }\!^{\lambda } + R_{\rho \nu \sigma }\!^{\lambda }
\xi _{\mu }\!^{\rho } + R_{\mu \rho \sigma }\!^{\lambda }\xi _{\nu }\!
^{\rho }\nonumber\\
+ R_{\mu \nu \rho }\!^{\lambda }\xi _{\sigma }\!^{\rho } -
R_{\mu \nu \sigma }\!^{\rho }\xi _{\rho }\!^{\lambda }\; , \label{eq:5} \\
\pounds_{\xi }T_{\mu \nu }\!^{\lambda } = \xi ^{\rho }\nabla _{\rho }
T_{\mu \nu }\!^{\lambda } + T_{\rho \nu }\!^{\lambda }\xi _{\mu }\!^{\rho }
+ T_{\mu \rho }\!^{\lambda }\xi _{\nu }\!^{\rho } -
T_{\mu \nu}\!^{\rho }\xi_{\rho }\!^{\lambda }\; . \label{eq:6}
\end{eqnarray}
\par
Comparing the results of the both Lie differentiation methods,
we get the covariant correlation, on the base of which, taking into
account that $\xi ^{\sigma }$ and $\nabla _{\lambda }\xi ^{\sigma }$ are
arbitrary, we obtain the two covariant identities. Because of the
arbitrariness of $\nabla _{\lambda }\xi ^{\sigma }$ we get
\begin{eqnarray}
\sqrt{-g}t_{\sigma }\!^{\lambda }[{\cal L}_{0}] = {\cal L}_{0}
\delta_{\sigma }\!^{\lambda } - 2R_{\mu\sigma \alpha }\!^{\beta }
\frac{\partial{\cal L}_{0}}{\partial R_{\mu \lambda \alpha }\!^{\beta }}
+ T_{\mu \nu }\!^{\lambda }\frac{\partial{\cal L}_{0}}
{\partial T_{\mu \nu}\!^{\sigma}}\nonumber\\
-2(\delta _{\sigma }\!^{\kappa }\hat {\nabla}_{\nu} -
T_{\sigma \nu }\!^{\kappa })\frac{\partial{\cal L}_{0}}
{\partial T_{\nu\lambda}\!^{\kappa}}\; , \label{eq:7}
\end{eqnarray}
where the following notations have been used:
\begin{eqnarray}
\sqrt{-g}t_{\sigma }\!^{\lambda }[{\cal L}_{0}]:=
\sqrt{-g}T_{\sigma }\!^{\lambda }[{\cal L}_{0}] + \hat {\nabla}_{\nu}
(\sqrt{-g}J_{\sigma }\!^{\lambda \nu}[{\cal L}_{0}])\; , \label{eq:8} \\
\sqrt{-g}T_{\sigma \rho }[{\cal L}_{0}]:= -2\frac{\delta{\cal L}_{0}}
{\delta g^{\sigma \rho}}\;, \;\;\;\;\;\;
\sqrt{-g}J_{\sigma }\!^{\lambda \nu}[{\cal L}_{0}]:=
-\frac{\delta{\cal L}_{0}}{\delta \Gamma_{\nu \lambda}\!^{\sigma}}\; .
\label{eq:9}
\end{eqnarray}
\par
Because of the arbitrariness of $\xi ^{\sigma }$ we get the second identity:
\begin{equation}
(\delta _{\sigma }\!^{\rho }\hat {\nabla}_{\lambda} -
T_{\sigma \lambda }\!^{\rho })\sqrt{-g}t_{\rho }\!^{\lambda }[{\cal L}_{0}]
- R_{\sigma \nu \lambda }\!^{\rho }
\frac{\delta{\cal L}_{0}}{\delta \Gamma_{\nu \lambda }\!^{\rho }}
- Q_{\sigma }\!^{\alpha \beta }\frac{\delta{\cal L}_{0}}
{\delta g^{\alpha \beta}} = 0\; . \label{eq:10}
\end{equation}
\par
When obtaining (\ref{eq:7}) and (\ref{eq:10}) the  following  identities  have
played the essential role
\begin{eqnarray}
\frac{\delta{\cal L}_{0}}{\delta \Gamma_{\sigma \nu}\!^{\lambda}} =
2\hat {\nabla}_{\rho}\frac{\partial{\cal L}_{0}}{\partial
R_{\sigma \rho \nu }\!^{\lambda }} + T_{\alpha \beta }\!^{\sigma }
\frac{\partial{\cal L}_{0}}{\partial R_{\alpha \beta \nu}\!^{\lambda}}
+ 2\frac{\partial{\cal L}_{0}}{\partial T_{\sigma\nu}\!^{\lambda}}\;,
\label{eq:11} \\
\hat {\nabla}_{\sigma}\frac{\delta{\cal L}_{0}}{\delta \Gamma_{\sigma
\nu}\!^{\lambda}} = (R_{\alpha \beta\sigma}\!^{\nu} \delta_{\lambda}\!^{\rho}
- R_{\alpha \beta\lambda}\!^{\rho} \delta_{\sigma}\!^{\nu})
\frac{\partial{\cal L}_{0}}{\partial R_{\alpha \beta \sigma}\!^{\rho}} +
2\hat {\nabla}_{\sigma}\frac{\partial{\cal L}_{0}}{\partial
T_{\sigma \nu}\!^{\lambda}}\; , \label{eq:12}
\end{eqnarray}
which take place in $(L_{4},g)$  for the the Lagrangian density (\ref{eq:3})
(\cite{Bab:Pon}, for some details see Appendix).
\par
Taking into account (\ref{eq:8}),(\ref{eq:9}), the field equations of the
metric-affine gravitational theory can be introduced in the following form
\begin{eqnarray}
\delta g^{\sigma \rho }: \qquad T_{\sigma \rho }[{\cal L}_{0}] =
- T_{\sigma \rho }\;, \qquad \qquad
T_{\sigma \rho }:= T_{\sigma \rho }[{\cal L}_{m}]\;, \label{eq:13} \\
\delta \Gamma _{\nu \lambda }\!^{\sigma }: \qquad J_{\sigma }\!^{\lambda\nu}
[{\cal L}_{0}] = - J_{\sigma }\!^{\lambda \nu }\; , \qquad
J_{\sigma }\!^{\lambda \nu }:= J_{\sigma }\!^{\lambda \nu }[{\cal L}_{m}]\;,
\label{eq:14}
\end{eqnarray}
where ${\cal L}_{m}$ is the Lagrangian density of matter, generating the
gravitational field. The system (\ref{eq:13}), (\ref{eq:14}) with the help of
the identity
\begin{equation}
R_{[\sigma \rho ]} = \frac{1}{2}\hat {\nabla}_{\tau }M_{\sigma \rho }
\!^{\tau }- \frac{1}{2}V_{\sigma \rho }\; , \qquad  M_{\sigma \rho }\!^{\tau}:
=T_{\sigma \rho }\!^{\tau } + 2\delta _{[\sigma ]}\!^{\tau }T_{\rho}\;,
\label{eq:15}
\end{equation}
where $V_{\alpha \beta }$ is the tensor of homothetic curvature of $(L_{4},g)$:
\begin{equation}
V_{\alpha \beta } = R_{\alpha \beta \tau }\!^{\tau } = \nabla _{[\alpha }
Q_{\beta]} + \frac{1}{2}T_{\alpha \beta }\!^{\tau }Q_{\tau }\; , \qquad
Q_{\sigma }:= Q_{\sigma \tau }\!^{\tau }\; , \label{eq:16}
\end{equation}
can be written in the equivalent form:
\begin{equation}
t_{\sigma }\!^{\lambda }[{\cal L}_{0}] = - t_{\sigma }\!^{\lambda}
\;,\qquad \qquad t_{\sigma }\!^{\lambda }:= t_{\sigma }\!^{\lambda }
[{\cal L}_{m}]\;. \label{eq:17}
\end{equation}
\begin{equation}
J_{\sigma }\!^{\lambda \nu}[{\cal L}_{0}] = - J_{\sigma }\!^{\lambda \nu}\;,
\label{eq:18}
\end{equation}
where in the right part of the gravitational field equations (\ref{eq:17}),
(\ref{eq:18}) we have the canonical energy-momentum tensor and the
hypermomentum
tensor of all matter that generates the gravitational field. These quantities
are calculated by means of the replacing the matter Lagrangian density
${\cal L}_{m}$ instead of ${\cal L}_{0}$ into (\ref{eq:8}),(\ref{eq:9}). The
question of deriving the field equations (\ref{eq:17}) will be discussed in
detail
in the following paper \cite{Bab:eq}.
\par
Substituting the field equations (\ref{eq:17}), (\ref{eq:18}) into
(\ref{eq:10}),
we obtain the quasiconservation law for the canonical  energy-momentum
tensor of matter in the metric-affine gravitational theory:
\begin{equation}
\hat {\nabla}_{\lambda}(\sqrt{-g}t_{\sigma }\!^{\lambda }) -
\sqrt{-g}T_{\sigma \lambda }\!^{\rho}t_{\rho}\!^{\lambda } +
\sqrt{-g}R_{\sigma \nu \lambda }\!^{\rho }J_{\rho }\!^{\lambda \nu } +
\frac{1}{2}\sqrt{-g}Q_{\sigma }\!^{\alpha \beta }T_{\alpha \beta } = 0 \;.
\label{eq:19}
\end{equation}
\par
The equation (\ref{eq:19}) was derived in \cite{He:fnd} (see also \cite{Hec})
as the consequence of the matter Lagrangian  invariance  with  respect  to
the infinitesimal coordinate transformations.
\vskip 0.5cm
\noindent
{\bf 3. The various cases of the matter motion in the metric-affine
       space-time $(L_{4},g)$}
\vskip 0.2cm
\par
The hypermomentum tensor can be split up in the following way \cite{He:nat}
\begin{equation}
J_{\sigma \rho }\!^{\lambda } = S_{\sigma \rho }\!^{\lambda } +
\overline{J}_{\sigma \rho }\!^{\lambda } + \frac{1}{4}g_{\sigma \rho }J^
{\lambda }\;, \qquad S_{\sigma \rho }\!^{\lambda }:= J_{[\sigma \rho ]}\!
^{\lambda }\;, \qquad \overline{J}_{\sigma }\!^{\sigma \lambda } = 0\;.
\label{eq:20}
\end{equation}
In $(L_{4},g)$ the affine connection coefficients have the form \cite{He:rev}
\begin{eqnarray}
\Gamma _{\lambda \nu }\!^{\sigma } = \frac{1}{2}g^{\sigma \rho }
\Delta ^{\alpha \beta \gamma }_{\nu \lambda\rho}(\partial_{\alpha}g_{\beta
\gamma}-T_{\alpha \beta \gamma } + Q_{\alpha \beta \gamma })\;, \label{eq:21}\\
\Delta^{\alpha\beta\gamma}_{\nu\lambda\rho}:= \delta^{\alpha }_{\nu }\delta
^{\beta}_{\lambda}\delta^{\gamma}_{\rho} + \delta^{\alpha}_{\lambda}\delta
^{\beta}_{\rho}\delta^{\gamma}_{\nu} - \delta^{\alpha}_{\rho}\delta^{\beta}
_{\nu}\delta^{\gamma}_{\lambda}\;.
\end{eqnarray}
\par
The quasiconservation law (\ref{eq:19}) with the help of (\ref{eq:20}),
(\ref{eq:21}) yields
\begin{eqnarray}
\sqrt{-g}\stackrel{(R)}{\nabla}_{\lambda}(t_{\sigma }\!^{\lambda }) +
\sqrt{-g}R_{\sigma \nu \lambda }\!^{\rho }S_{\rho }\!^{\lambda \nu } +
\frac{1}{4}\sqrt{-g}V_{\sigma \nu }J^{\nu } \nonumber \\
- (T_{\sigma [\lambda }\!^{\rho ]} +
\frac{1}{2}T^{\rho}\!_{ \lambda \sigma} - Q^{[\rho}\!_{\lambda ] \sigma} +
\frac{1}{2}Q_{\sigma \lambda }\!^{\rho })\hat {\nabla}_{\nu}
(\sqrt{-g}J_{\rho }\!^{\lambda \nu })  \nonumber \\
+ \sqrt{-g}(\nabla _{[\sigma }Q_{\nu ] \lambda \rho } +
\frac{1}{2}T_{\sigma \nu }\!^{\tau}Q_{\tau \lambda \rho })
\overline{J}^{\lambda \rho \nu} = 0\;. \label{eq:22}
\end{eqnarray}
\par
Let us consider some particular cases of the matter motion.
\par
I. The matter hypermomentum tensor is equal to the spin momentum
tensor: $J_{\sigma \rho }\!^{\lambda } = S_{\sigma \rho }\!^{\lambda }$.
\par
In this case some terms with nonmetricity vanish  in  (\ref{eq:22})  exept
the terms $(Q^{[\rho}\!_{\lambda]\sigma}-\frac{1}{2}Q_{\sigma\lambda}\!
^{\rho})\hat {\nabla}_{\nu}(\sqrt{-g}S_{\rho}\!^{\lambda\nu})\;$.
Therefore in this  case  the matter motion depends on both torsion
and nonmetricity.
\par
II. The matter hypermomentum tensor vanishes: $J_{\sigma \rho }\!^{\lambda } =
0\;$.
\par
In this case the canonical energy-momentum tensor of  matter (\ref{eq:17})
reduces to the metric one: $t_{\sigma \lambda } = T_{\sigma \lambda }$ and
the quasiconservation  law (\ref{eq:22}) reduces to the matter energy-momentum
conservation  law  in  a Riemann space-time (\ref{eq:1}). Therefore we have
proved the following Theorem \cite{Bab:NC,Bab:Tmsk}.
\par
{\sl Theorem}: In $(L_{4},g)$ the motion of matter without
hypermomentum coinsides with the motion in the Riemann space-time,
which the metric tensor coinsides with the metric tensor of $(L_{4},g)$.
\par
Thus bodies and mediums without hypermomentum are  not  subjected
to the influence of the possible nonmetricity of the space-time (in
contrast to the generally accepted opinion) and therefore  can  not
be the tools for  the  detection  of  the  deviation  of  the  real
space-time properties from the Riemann space structure.
\par
Therefore for the investigation of the  different  manifestations
of the possible space-time nonmetricity one needs to use the bodies
and  mediums  endowed  with  the  hypermomentum,  i.e.   the   spin
particles, the perfect spinning fluid or the perfect fluid with the
intrinsic hypermomentum \cite{Bab:UDN,Bab:Arg,Bab:MPGU}.
\vskip 0.5cm
\noindent
{\bf Appendix}
\vskip 0.2cm
\par
        Let us consider the identities (\ref{eq:11}),(\ref{eq:12}) in more
detail. The identity (\ref{eq:11}) is the direct consequence of (\ref{eq:3})
and the explicit forms of the curvature tensor $R_{\mu \nu \sigma}\!^{\lambda
}$
and the torsion tensor $T_{\mu \nu}\!^{\lambda}$.
\par
        In order to derive the identity (\ref{eq:12}) let us evaluate  the
covariant derivative $\hat {\nabla}_{\sigma}$ from the both sides of
the (\ref{eq:11}):
\begin{eqnarray}
\hat {\nabla}_{\sigma}\frac{\delta{\cal L}_{0}}{\delta
\Gamma_{\sigma \nu}\!^{\lambda}} =
2\nabla_{[\sigma}\nabla_{\alpha]}\frac{\partial{\cal L}_{0}}{\partial
R_{\sigma \alpha\nu }\!^{\lambda }} + 2(\nabla_{\sigma}T_{\alpha})
\frac{\partial{\cal L}_{0}}{\partial R_{\sigma\alpha\nu}\!^{\lambda}}
\nonumber \\
+\nabla_{\sigma}\left(T_{\alpha\beta}\!^{\sigma}\frac{\partial{\cal L}_{0}}
{\partial R_{\alpha\beta\nu}\!^{\lambda}}\right) +
T_{\sigma}T_{\alpha\beta}\!^{\sigma}\frac{\partial{\cal L}_{0}}
{\partial R_{\alpha\beta\nu}\!^{\lambda}} +
2\hat {\nabla}_{\sigma}\frac{\partial{\cal L}_{0}}{\partial
T_{\sigma \nu}\!^{\lambda}}\;.\label{eq:23}
\end{eqnarray}
The first term in the right part of the (\ref{eq:23}) may be calculated with
the help of Ricci identity in $(L_4,g)$  \cite{Sch}:
\begin{equation}
2\nabla_{[\alpha}\nabla_{\beta]}{\cal W}^{\lambda} = R_{\alpha\beta\sigma}\!
^{\lambda}{\cal W}^{\sigma} - T_{\alpha\beta}\!^{\sigma}
\nabla_{\sigma}{\cal W}^{\lambda}
- V_{\alpha\beta}{\cal W}^{\lambda}\;. \label{eq:24}
\end{equation}
Here $V_{\alpha\beta}$ is the tensor of homothetic curvature
(\ref{eq:16}); ${\cal W}^{\lambda}$ is a contravariant vector
density, the minus  appearing in the first term in the right part of the
(\ref{eq:24}) in case of the covariant vector density ${\cal W}_{\lambda}$.
As a result one has
\begin{eqnarray}
\hat {\nabla}_{\sigma}\frac{\delta{\cal L}_{0}}{\delta
\Gamma_{\sigma \nu}\!^{\lambda}} = (2\nabla_{\alpha}T_{\beta}
+ \nabla_{\sigma}T_{\alpha\beta}\!^{\sigma} + T_{\sigma}T_{\alpha\beta}\!
^{\sigma} - 3R_{[\alpha\beta\sigma]}\!^{\sigma})\frac{\partial{\cal L}_{0}}
{\partial R_{\alpha\beta\nu }\!^{\lambda }} \nonumber \\
+R_{\alpha\beta\sigma}\!^{\nu}\frac{\partial{\cal L}_{0}}
{\partial R_{\alpha\beta\sigma}\!^{\lambda}} -
R_{\alpha\beta\lambda}\!^{\sigma}\frac{\partial{\cal L}_{0}}{\partial
R_{\alpha\beta\nu}\!^{\sigma}} +
2\hat {\nabla}_{\sigma}\frac{\partial{\cal L}_{0}}{\partial
T_{\sigma \nu}\!^{\lambda}}\;.\label{eq:25}
\end{eqnarray}
Then using in (\ref{eq:25}) {\em the second identity for the curvature}
\cite{Sch}:
\begin{equation}
R_{[\alpha\beta\sigma]}\!^{\lambda} = \nabla_{[\alpha}T_{\beta\sigma]}\!
^{\lambda}-T_{[\alpha\beta}\!^{\rho}T_{\sigma]\rho}\!^{\lambda}\;,
\label{eq:26}
\end{equation}
we convince oneself that the terms in the parentheses in the right part of
the (\ref{eq:25}) vanishes. As a result we get the identity (\ref{eq:12}).

\end{document}